\documentclass[twocolumn,showpacs,preprintnumbers,amsmath,amssymb]{revtex4}

\usepackage{graphicx}
\usepackage{dcolumn}
\usepackage{bm}

\begin{document}


\title{Charge Spin Separation in 3D}

\author{M. Cristina Diamantini}
\email{cristina.diamantini@pg.infn.it}
\affiliation{%
INFN and Dipartimento di Fisica, University of Perugia, via A. Pascoli, I-06100 Perugia, Italy
}%


\author{Carlo A. Trugenberger}
\email{ca.trugenberger@InfoCodex.com}
\affiliation{%
SwissScientific, chemin Diodati 10, CH-1223 Cologny, Switzerland
}%


\date{\today}

\begin{abstract}
Electron fractionalization into spinons and chargeons plays a crucial role in 2D models of strongly correlated electrons. In this paper we show that spin-charge separation is not a phenomenon confined to lower dimensions but, rather, we present a field-theoretic model in which it is realized in 3D. The model involves two gauge fields, a standard one and a two-form gauge field. The physical picture is that of a two-fluid model of chargeons and spinons interacting by the topological BF term. When a Higgs mechanism of the second kind for the two-form gauge field takes place, chargeons and spinons are bound together into a charge 1 particle with spin 1/2. The mechanism is the same one that gives spin to quarks bound into mesons in non-critical string theories and involves the self-intersection number of surfaces in 4D space-time. A state with free chargeons and spinons is a topological insulator. When chargeons condense, the system becomes a topological superconductor; a condensate of spinons, instead realizes $U(1)$ charge confinement. 
\end{abstract}
\maketitle

\section{Introduction}
The concept of spin-charge separation is one of the guiding principles of the modern approach to strongly correlated, low-dimensional systems \cite{ave}. The idea is that, in specific ground states, the electron is fractionalized into two "constituent" quasi-particles, the chargeon (holon) carrying only the charge degree of freedom and the spinon, carrying only the spin degree of freedom. The two quasi-particles interact via emergent gauge fields: the electron is reconstituted when the gauge interaction becomes strong enough to cause confinement. 

The idea originates from the work of Tomonaga and Luttinger and was shown by Haldane \cite{hal} to be a generic feature of 1D metallic systems. Moreover, the idea of electron fractionalization is thought to play a crucial role in the physics of the high-$T_c$ cuprates. Indeed, spin-charge separation seems to be an unavoidable characteristics of the 2D $tJ$ model \cite{weng, ye} of the doped Mott insulators, capturing the essential physics of high-$T_c$ superconductivity \cite{wensach}. There is an abundance of evidence that electron fractionalization in these models leads to new quantum orders not characterized by symmetry breaking \cite{sen1}. 

A key ingredient of the spin-charge separation idea in 2D is the representation of chargeons and spinons as a two-fluid model with mutual Chern-Simons interactions \cite{wensach}, a picture that can be analytically derived from the $tJ$ model \cite{weng, ye}. Mixed Chern-Simons fluids as representations of condensed matter systems where first introduced in \cite{dst1}, where it was shown that they capture all the essential physics of 2D Josephson junction arrays. In the same paper it was also pointed out that this two-fluid construction can be generalized to 3D, the topological interaction being encoded in what is known as the $BF$ term \cite{birmi}.

Based on this construction we proposed a new, topological mechanism for superconductivity based on the condensation of topological defects in a BF model \cite{dst2}, no symmetry breaking being involved. A pure BF theory has also been recently shown \cite{moo} to represent the long-distance physics of 3D topological insulators \cite{topins}. The electric permittivity and magnetic permeability of the material govern quantum phase transitions from topological insulators to topological superconductors and possibly to a $U(1)$ charge confinement phase \cite{dst3}. 

Spin- charge separation is mostly believed to be a phenomenon confined to lower space dimensions (1D and 2D), where it is typically derived from fermionic models like, e.g., the Hubbard model. In this paper we show that this is not so. The topological excitations driving the topological insulator-topological superconductor and topological insulator-confinement quantum phase transitions in 3D are indeed chargeons and spinons. We show that the presence of a specific combination of relevant terms in the action leads to a generalized Higgs mechanism of the second kind for the two-form gauge field. When this happens, chargeons and spinons are bound together and the resulting excitation describes a charge 1 particle with spin 1/2. The spin arises in a subtle way: particle-antiparticle fluctuations about the symmetry-broken ground state are connected by a string whose only action is a topological term measuring the self-intersection number of the world-surface it sweeps in 4D (Euclidean) space-time. As was shown in \cite{paw}, this factor is equivalent to the spin factor for fermionic particles in the path integral formalism \cite{pol1}. Indeed, this is the same mechanism giving spin to the quarks bound together in mesons in non-critical string theories, the only difference being that in the present case the string tensions and all other curvature terms vanish, so that the particles are not confined and the string carries only the spin information. This way we derive the fermionization of a bosonic two-fluid model rather then the other way around.

The idea is to represent chargeons and spinons as topological excitations of charge and spin gauge fields. These topological excitations represent quasi-particles that arise due to the compactness of the corresponding gauge groups, the charge and spin fields mediating the emergent gauge interactions between these quasi-particles.

A massive spin 1/2 particle in 3D is characterized by three degrees of freedom, a scalar charge degree of freedom and two degrees of freedom for the spin. Representing the spin as a separate bosonic entity, the spin field, requires thus a massive vector with two degrees of freedom. It is well known that, in 3D, massless vectors (photons) carry two degrees of freedom, called helicities, while massive vectors obtained from spontaneous symmetry breaking and described by the Proca Lagrangian carry three degrees of freedom. This is the reason why it is mostly believed that spin-charge separation is impossible in 3D. It is, however not widely appreciated that there is a way to describe a vector particle in 3D with a gauge invariant mass. Exactly as its Chern- Simons counterpart in 2D \cite{jackiw}, the gauge invariant mass arises from a topological term in the action \cite{bow}: such vector particles in 3D carry thus only two helicity degrees of freedom and are ideal candidates to describe spin fields.  

Spin fields are thus described by a vector particle $a_{\mu}$ with gauge invariance under the transformations:
\begin{equation}
a_{\mu} \to a_{\mu} + \partial _{\mu} \lambda \ .
\label{one}
\end{equation}
Since $a_0$ plays the role of a non-dynamical Lagrange multiplier in the action, this gauge invariance eliminates one degree of freedom from the remaining three, leaving the two helicities as the only surviving degrees of freedom. 

Charge fields must be described by a single scalar degree of freedom. There are two ways to describe a massive scalar in 3D, either directly or by embedding it in an antisymmetric gauge potential of the second kind $b_{\mu \nu}$ with gauge invariance under the transformations:
\begin{equation}
b_{\mu \nu} \to b_{\mu \nu} + \partial_{\mu} \eta_{\nu} - \partial_{\nu} \eta_{\mu} \ .
\label{two}
\end{equation}
The mixed components $b_{0i}=-b_{i0}$ will also play the role of non-dynamical Lagrange multipliers, leaving three dynamical degrees of freedom. The above gauge invariance eliminates, however, two of these, since there are two independent gauge parameters $\eta_i$ (the other one being eliminated by the equivalence $\eta_i\equiv \eta_i+ \partial _i \rho $), leaving thus one overall degree of freedom. 

We will now consider low-energy effective theories for condensed matter system which are formulated in terms of the charge and spin field degrees of freedom just introduced. The most general such model, containing only relevant and marginal terms and compatible with the two separate gauge invariances (\ref{one},\ref{two}) is given by:
\begin{eqnarray}
S &&= {ik \over 4 \pi} \int d^4x \ b_{\mu \nu} \epsilon_{\mu \nu \alpha \beta} f_{\alpha \beta} +
{i\theta \over 16 \pi^2} \int d^4x \ f_{\mu \nu} \tilde f_{\mu \nu} +
\nonumber \\
&&+  {1\over 4 e^2 } \int d^4x \ f_{\mu \nu} f_{\mu \nu}  \ ,
\label{three}
\end{eqnarray}
where $f_{\mu \nu} = \partial_{\mu}a_{\nu}-\partial_{\nu}a_{\mu}$ is the field strength associated with $a_{\mu}$, $\tilde f_{\mu \nu} = (1/2) \epsilon_{\mu \nu \alpha \beta} f_{\alpha \beta}$ its dual
and $k$ (the BF coupling), $\theta$ and $e$ are dimensionless couplings. 
We use relativistic notation in Euclidean space-time: possible non-relativistic effects would not alter the main conclusions.

The first two terms in this action are purely topological terms: the first is called generically the $BF$ term \cite{birmi} and represents a generalization to 3D of the mutual Chern-Simons terms in 2D. It preserves the $P$ amd $T$ symmetries if the two-form gauge field is a pseudotensor.
The second is the famed $\theta$-term of axion electrodynamics \cite{wil}. The parameter $\theta $ is an angle variable with periodicity $2\pi$, the partition function being invariant under the shift $\theta \to \theta +2\pi$. The $\theta$-term breaks generically the $P$ and $T$ symmetries: these are however restored when $\theta$ is quantized: $\theta = n \pi$, $n \in \mathbb{ Z}$. Thus there are only two possible $\theta $ values compatible with the $P$ and $T$ symmetries: $\theta = 0$ and $\theta = \pi$. The third term in the action (\ref{three}) has the form of a standard Maxwell term for the effective gauge field $a_{\mu}$. 

The physical interpretation is that 
\begin{equation}
j_{\mu} = {k\over 2\pi} \epsilon_{\mu \nu \alpha \beta}\partial_{\nu}b_{\alpha \beta} \    , 
\phi_{\mu \nu} = {k\over 2\pi} \epsilon_{\mu \nu \alpha \beta}\partial_{\alpha} a_{\beta} \ ,
\label{four}
\end{equation}
are the conserved charge and spin currents representing the low-energy fluctuations about a topologically ordered state. When the two Abelian gauge symmetries (\ref{one},\ref{two})  are $U(1)$ compact symmetries, the dual field strengths (\ref{four}) contain singularities \cite{pol2} 
\begin{eqnarray}
J_{\mu} &&= \int_C d\tau \ {dx_{\mu}(\tau)\over d\tau} \ \delta^4 \left( {\bf x} - {\bf x(\tau)} \right) \ ,
\nonumber \\
\Phi_{\mu \nu} &&= {1\over 2} \int _S d^2 \sigma \ X_{\mu \nu} ({\sigma}) \ \delta^4 \left( {\bf x} - {\bf x(\sigma)} \right) 
\nonumber \\
X_{\mu \nu} &&= \epsilon^{ab} {\partial x_{\mu} \over \partial \sigma^a} {\partial x_{\nu} \over \partial \sigma^b} \ ,
\label{five}
\end{eqnarray}
where $C$ and $S$ are closed curves and compact surfaces parametrized by ${\bf x(\tau)}$ and ${\bf x(\sigma)}$ respectively. These have standard couplings to the charge and spin gauge fields: $ ik a_{\mu} J_{\mu}$ and $ikb_{\mu \nu} \Phi_{\mu \nu}$ respectively 
and describe chargeon and spinon quasi-particle fluctuations about the ground state. Since in 3D the spinon is a vector with two degrees of freedom, the allowed polarizations are transverse when it is moving, exactly as in the case of a standard photon. Since, however, it is a massive vector particle, it can point in any space direction in its rest frame. So, in this case, it is the direction of movement rather than the polarization that is restricted: a spinon is a massive vector particle that moves always perpendicularly to the direction in which it is pointing. It has thus "quantum Hall-type" responses to external fields. 

In order to make computations tractable we shall add, as a regulator, an infrared-irrelevant but gauge invariant kinetic term for the charge degree of freedom,
\begin{equation}
S\to S+S_{\rm reg}, \ S_{\rm reg} = {1\over 12 \Lambda^2} \int d^4 x \ h_{\mu \nu \alpha}h_{\mu \nu \alpha} \  ,
\label{seven}
\end{equation}
where $h_{\mu \nu \alpha} =  \partial_\mu b_{\nu \alpha} + \partial_\nu b_{ \alpha \mu} +\partial_\alpha b_{\mu \nu }$ is the field strength associated with the two-form gauge field $b_{\mu \nu}$ and $\Lambda$ is a mass parameter of the order of the ultraviolet cutoff $\Lambda_0$. This regulator term makes the quadratic kernels well defined by inducing a mass $m=e\Lambda k/\pi$ for all fields. This is the anticipated topological, gauge invariant mass \cite{bow} that is the 3D analogue of the famed Chern-Simons topological mass \cite{jackiw}. This mass represents the gap for the topologically-ordered state: it sets the energy scale for charge- and spin-wave excitations and the thickness scale for chargeon and spinon quasi-particle excitations. It thus plays the same role as the inverse magnetic length in the quantum Hall effect. The mass can be removed again after integrations by letting $\Lambda \to \infty$: in this case the gap becomes infinite and only point-like chargeons and spinons quasi-particles survive. 

It is easy to establish that the model (\ref{three}) describes a topological insulator \cite{topins}. First of all, the charge degrees of freedom, carried by the two-form gauge field $b_{\mu \nu}$, have no dynamics in the bulk of the material since the $BF$ term is a topological term. The only dynamic matter degrees of freedom are edge modes describing surface Dirac fermions \cite{moo}. Secondly, let us consider the coupling of the matter currents (\ref{four}) to an external electromagnetic field $A_{\mu}$ dictated by the form of the effective action (\ref{three}), 
\begin{equation}
S \to S + i \int d^4 x   j_{\mu}A_{\mu}  \ .
\label{eight}
\end{equation}
In presence of a non-vanishing angle $\theta$, the spin field carries vorticity. The $BF$ coupling (contained in the definition of the current $j_{\mu}$) is the charge unit of chargeons whereas $\theta$ is the flux unit of spinons. 
Integrating out all matter degrees of freedom one obtains, in the limit $\Lambda \to \infty$, the effective electromagnetic action
\begin{equation}
S_{\it eff} = \int d^4x  \ {1 \over 4e^2} \ F_{\mu \nu}F_{\mu \nu} + 
{i  \theta  \over 16 \pi^2} F_{\mu \nu}\tilde F^{\mu \nu}  \ .
\label{eight}
\end{equation}
This describes a bulk dielectric material with an axion electrodynamics term that characterizes strong topological insulators when $\theta = \pi$. 

In \cite{dst3} we have shown that a condensation of chargeons turns the topological insulator into a topological superconductor, while the condensation of spinons leads to a $U(1)$ charge confinement regime. In what follows we will concentrate on a different quantum phase transition.

It is well known that adding marginal terms to an action can drive the system to a new fixed point, describing an entirely different physics. In the present case there are indeed three additional marginal terms that can be added to the model (\ref{three}): $b_{\mu \nu} b_{\mu \nu} $, $b_{\mu \nu} f_{\mu \nu}$ and $b_{\mu \nu} \epsilon _{\mu \nu \alpha \beta } b_{\alpha \beta} $. All these terms, taken one by one, break the gauge invariance (\ref{two}) and introduce thus new, unwanted degrees of freedom. There is, however one particular combination of these three terms that can be added to the effective action and that preserves both gauge invariances, albeit (\ref{two}) is realized in a different, more subtle way:
\begin{eqnarray}
S &&= {ik^2 \over 2 \theta} \int d^4x \ \left( b_{\mu \nu}+{\theta \over 4\pi k}f_{\mu \nu} \right) 
\epsilon_{\mu \nu \alpha \beta} \left( b_{\alpha \beta} +{\theta \over 4\pi k} f_{\alpha \beta} \right) 
\nonumber \\
&&+  {4 \pi^2k^2 \over  e^2\theta^2 } \int d^4x \ \left( b_{\mu \nu} + {\theta \over 4\pi k}f_{\mu \nu}\right) \left( b_{\mu \nu} + {\theta \over 4\pi k}f_{\mu \nu} \right) ,
\label{nine}
\end{eqnarray}

Something very interesting happens when the additional marginal terms in the effective action combine with the original ones to give (\ref{nine}). Only the combination $\left( b_{\mu \nu}+{\theta \over 4\pi k}f_{\mu \nu} \right)$ appears in the action and the tensor gauge invariance (\ref{two}) is preserved if it is combined with a corresponding shift  $a_{\mu } \to a_{\mu } - {4\pi k\over \theta} \eta_{\mu }$. This combined transformation can be exploited to entirely absorb $f_{\mu \nu}$ into $b_{\mu \nu}$, giving the effective action
\begin{eqnarray}
S &&= {i \over 32 \theta} \int d^4x \ b_{\mu \nu} \epsilon_{\mu \nu \alpha \beta} b_{\alpha \beta} 
+  \int d^4 x {\pi^2 \over  4 e^2\theta^2 } b_{\mu \nu} b_{\mu \nu}  
\nonumber \\
&&+ {i\over 16\pi} \int d^4x \ b_{\mu \nu} \epsilon_{\mu \nu \alpha \beta} F_{\alpha \beta} + i \int d^4x  \ b_{\mu \nu} \Phi_{\mu \nu} ,
\label{ten}
\end{eqnarray}
where we have rescaled, for notational simplicity, the tensor gauge field by a factor 4 and we have included the couplings to external electromagnetic fields and to quasi-particle excitations. Note that the original $BF$ coupling $k$ falls completely out of the action and is replaced by the factor $\theta / \pi$, that takes over the role of charge unit.  

In this gauge-fixed form, the original gauge symmetry (\ref{two}) appears as broken. This is nothing else than a Higgs mechanism of the second kind for the tensor gauge symmetry ($\ref{two}$). This is also known as the St\"uckelberg mechanism \cite{stu}, in which a scalar longitudinal polarization "eats up" two transverse polarizations to become a massive vector. The St\"uckelberg mechanism is the dual of the standard Higgs mechanism and is thus responsible for confinement \cite{stu}, soldering in this case chargeons to spinons. 

Indeed, corresponding to the merger of the spin and charge gauge fields $a_{\mu }$ and $b_{\mu \nu }$ into a single tensor field with 3 massive degrees of freedom, there is also a merger of chargeons and spinons into a unique string-like quasi-particle excitation with open world-sheets, describing 
magnetic fluxes with charged dyons at their ends. The closed boundaries of the open surfaces represent the world-lines of dyon-antidyon fluctuations with charge $\theta / \pi$ and current $J_{\mu } = (1/2\pi) \partial_{\nu}\Phi_{\mu \nu}$ as can be inferred from the induced electromagnetic action 
\begin{eqnarray}
S_{\it eff} = \int d^4x  \ {1 \over 4e^2} \ F_{\mu \nu}F_{\mu \nu} + 
{i  \theta  \over 16 \pi^2} F_{\mu \nu}\tilde F^{\mu \nu} 
\nonumber \\
+ i \int d^4 x \ {(\theta /\pi) \over 2\pi} A_{\mu} \partial_{\nu}\Phi_{\mu \nu} + {4\pi \over e^2} \tilde F_{\mu \nu} \Phi_{\mu \nu} \ .
\label{eleven}
\end{eqnarray}
This is the same induced action as in (\ref{nine}) but the matter degrees of freedom have now dynamics in the bulk. 

In order to gain more insight into the character of the resulting soldered quasi-particle let us compute its induced action by using the explicit form (\ref{five}) for $\Phi_{\mu \nu}$. The relevant and marginal terms in this action are
\begin{eqnarray}
&&S_{QP} = {\Lambda^2 \over 4\pi} K_0 \left( {m\over \Lambda_0} \right) \int_S d^2\sigma \sqrt{g} 
+ {\Lambda^2\over 16\pi m^2} \int _S d^2 \sigma \sqrt{g} R 
\nonumber \\
&& - {\Lambda^2\over 16\pi m^2} \int_S d^2 \sigma \sqrt{g} g^{ab} \partial_a t_{\mu \nu} \partial_b t_{\mu \nu}
\nonumber \\
&&-i{\theta \over \pi} {\pi \over {1+{4\pi \over e^2}}} \nu + {e^2 m\over 8\pi^2} f \left( {m\over \Lambda_0} \right)
\int_{\it \partial S} d\tau \sqrt{ {dx_{\mu}\over d\tau} {dx_{\mu}\over d\tau}} \ ,
\label{twelve}
\end{eqnarray}
where $m = e\Lambda/4\pi$, $f(x) = \int_x^{\infty} dz K_1(z)/z$ and $K_0$ and $K_1$ are Bessel functions of imaginary argument, with asymptotic behaviours
$K_0(x) \simeq {\rm exp}(-x)/\sqrt{x}$ and $f(x) \simeq {\rm exp}(-x)/\sqrt{x^3}$ for large x. 
The geometric quantities in this expression are defined in terms of the induced surface metric $g_{ab} = {\partial x_{\mu}\over \partial \sigma ^a} {\partial x_{\mu }\over \partial \sigma ^b}$ as $g = {\rm det}\  g_{ab} = X_{\mu \nu} X_{\mu \nu}/2$ and $t_{\mu \nu} =  X_{\mu \nu}/ \sqrt{g}$. The quantity $R$ is the scalar (intrinsic) curvature of the world-surface while 
\begin{equation}
\nu = {1\over 4\pi} \int d^2\sigma \ \sqrt{g} \epsilon_{\mu \nu \alpha \beta} g^{ab} \partial_a t_{\mu \nu}
\partial_b t_{\alpha \beta} \ ,
\label{thirteen}
\end{equation}
represents the (signed) self-intersection number of the world-surface. 

The important point is that, to avoid infinities and obtain a well-defined boundary term when we remove the UV cutoff, the renormalization flow implies $e\to \infty$ as $\Lambda_0 \to \infty$ (and $\Lambda = {\rm const.} \Lambda_0 \to \infty $). As a consequence of this running coupling constant, both the string tension and the curvature terms in (\ref{thirteen}) vanish at large distances and the induced quasi-particle action becomes 
\begin{equation}
S_{QP} = m_s \int_{\it \partial S} d\tau \sqrt{ {dx_{\mu}\over d\tau} {dx_{\mu}\over d\tau}} -i{\theta \over \pi} \pi \nu \ ,
\label{fourteen}
\end{equation}
where $m_s$ is the renormalized mass of the particle. Note that the only remnant of the string at large distances is the topological self-intersection number. It has been shown \cite{paw} that at $\theta /\pi = 1$ this topological term is just another representation of the spin factor of a point particle with spin 1/2. This shows that, in the Higgs phase (of the second kind) chargeons and spinons recombine into a single particle with charge 1 and spin 1/2. 

We thus conclude that spin-charge separation can occur in 3D when a non trivial axion term with $\theta$-angle $\pi $ is generated in the electromagnetic action. The phase with free separated chargeons and spinons is a strong topological insulator. Topological superconductivity occurs when chargeons condense and an exotic $U(1)$ confinement would be realized in presence of spinon condensate.

\end{document}